\documentclass[usenatbib]{mnras}
\usepackage{xcolor}
\usepackage{hyperref}
\usepackage{enumerate}
\usepackage [caption=false]{subfig}
\usepackage{natbib}
\usepackage{url}
\usepackage{graphicx}	
\usepackage{amsmath}	
\usepackage{amssymb}
\usepackage{mathrsfs}
\newcommand\Nu{NuSTAR }
\newcommand\ec{$E_{\rm cut} $ }

\newcommand\first{Bias\uppercase\expandafter{\romannumeral1 } }
\newcommand\second{Bias\uppercase\expandafter{\romannumeral2 } }

\title[Biases in flux-resolved spectroscopy]{Hidden biases in flux-resolved X-ray spectroscopy}

\author[J.-L. Kang and J.-X. Wang]{
Jia-Lai Kang$^{1,2}$\thanks{ericofk@mail.ustc.edu.cn}
and Jun-Xian Wang$^{1,2}$\thanks{jxw@ustc.edu.cn}
\\
$^{1}$CAS Key Laboratory for Research in Galaxies and Cosmology, Department of Astronomy, University of Science and Technology of China, \\
  Hefei, Anhui 230026, China\\
$^{2}$School of Astronomy and Space Science, University of Science and Technology of China, Hefei 230026, China\
}

\pubyear{2022}

\begin{document}
\label{firstpage}
\pagerange{\pageref{firstpage}--\pageref{lastpage}}
\maketitle
\begin{abstract}
Flux-resolved X-ray spectroscopy is widely adopted to investigate the spectral variation of a target between various flux levels. In many cases it is done through horizontally splitting a single light curve into multiple flux levels with certain count rate threshold(s). In this work we point out there are two hidden biases in this approach which could affect the spectral analyses under particular circumstances. The first is that, when Poisson fluctuations of the source counts in light curve bins are non-negligible compared with the intrinsic variation, this approach would over-estimate (under-estimate) the intrinsic average flux level of the high (low) state. The second bias is that, when the Poisson fluctuations of the background count rate is non-negligible, the background spectrum of the high (low) state would be under-estimated (over-estimated), thus yielding biased spectral fitting parameters. We take NuSTAR spectra for example to illustrate the effects of the biases, and particularly, how the measurements of the coronal temperature in AGNs would be biased. We present a toy method to assess the significance of such biases, and approaches to correct for them when necessary. 
\end{abstract}


\section{Introduction} \label{sec:intro}
\par Flux-resolved X-ray spectroscopy is a widely used technique to investigate the spectral variation of individual X-ray sources between different flux levels. Generally, it could be conducted through splitting a light curve either vertically (with vertical cuts in time) into several consecutive segments \citep[e.g.,][]{Keek2016, Shidatsu2017, Turner2018, Buisson_2021}, or horizontally (with cuts in count rate) into fragments with various flux levels \citep[e.g.][]{Vaughan2004, Zoghbi2010, Parker2014, Parker2017, Barua2020, Barua2021,Jiang_2022}. While the former method is practically no different from studying the spectral variations between multiple observations, the latter is also useful in cases when it is hard to split a light curve vertically into a small number of segments with significantly distinct flux levels (for example, if the variability in a light curve is dominated by high frequency variations).

\par In this work we point out that, there are two hidden biases unnoticed previously in this latter approach, i.e., splitting a light curve horizontally with cuts in net count rate. In principle, the splitting could be done with more than one cut in the count rate. For instance, one may split the light curve into three flux levels (i.e., low, intermediate and high) with two cuts in the count rate. In this work, for simplicity, we only consider the simplest case with a single count rate cut, i.e., splitting a single light curve into low and high states. However, the concepts of the biases and the corrections we present in this work are also applicable to more general cases.

\par The first bias is that, when Poisson fluctuations of the source counts in the light curve bins are non-negligible compared with the intrinsic variation, this horizontal splitting approach would over-estimate (under-estimate) the intrinsic average flux of the corresponding high (low) state spectrum. This is because some light curve bins could have been promoted into the high state (or demoted into the low state) solely because of Poisson fluctuations. The second but more important bias is that, when the Poisson fluctuations of background counts are non-negligible, similarly, the background spectrum of the high (low) state would be under-estimated (over-estimated), thus yielding biased spectral fitting parameters. Hereafter these two biases are referred to as \first and \second respectively. Note the Poisson fluctuations of source photons would not cause under-/over-estimation of background, thus have no contribution to Bias II. Meanwhile, the Poisson fluctuations of background photons do yield biased net count rates (source - background), which is however included in Bias II, thus independent to BiasI. The significance of these two biases depends on complicated factors, including the brightness of the source, the background fraction, the intrinsic variability amplitude, the bin size of the light curves, etc. 

\par While the biases could in principle affect data obtained with all relevant instruments, in this paper we take simulated NuSTAR data for example to illustrate the effects of such biases. We also present a simple method to assess the significance of such biases, and approaches to correct for them when necessary. The Nuclear Spectroscopic Telescope Array \citep[NuSTAR, ][]{Harrison_2013}, the first focusing high-energy X-ray mission with a broad spectral coverage from 3 to 79 keV, has enabled the constraints of high-energy features in the spectra. A much concerned parameter is the high-energy cutoff $E_{\rm cut}$, a direct indicator of the coronal temperature $T_{\rm c}$ in AGNs \citep[e.g.][]{Tortosa_2018, Rani_2019, Kang_2020, Panagiotou_2020, Hinkle_2021, Kang_2022} and X-ray binaries \citep[e.g.][]{Motta_2009, Yan_2020, Mendez_2022}. Investigating the variation of $E_{\rm cut}$/$T_{\rm c}$ could reveal the dynamic nature of the corona \citep[e.g.][]{Ballantyne_2014, Ursini_2016, Zhangjx2018, Kang_2021, Wilkins_2022, Pal_2022a, Ding_2022, Pal_2022b}. Accurate measurements of the $T_{\rm c}$/$E_{\rm cut}$ are extremely sensitive to the high-energy data where the background could dominate. It is thus essential to explore whether the aforementioned \second could significantly alter the measured $T_{\rm c}$/$E_{\rm cut}$.  

\section{Methods}\label{sec:method}
\par In this work we utilize simulated NuSTAR data with different input spectral parameters to illustrate the effects of the biases under different conditions. We assume the NuSTAR source spectra and light curves are extracted in a region with $\rm Area_{src}$ (typically a circle with a radius of 60\arcsec), background from a source free region with $\rm Area_{bkg}$,  and $\rm f_{scale}$ defined as the ratio of $\rm Area_{src}$ to $\rm Area_{bkg}$. For simplicity we adopt $\rm f_{scale}$ = 1.0 in our simulations below.  Adopting smaller $\rm f_{scale}$ (relatively larger region to extract the background light curve and spectrum) could reduce, but not erase, the effect of BiasII, as even if the intrinsic background level could be precisely measured within an infinitely large background region, the background counts in the source region could still fluctuate randomly due to Poisson noises. 

\par For each group of input parameters, the simulation is performed as following. First we simulate a source light curve with certain variation amplitude and power density spectrum, using pyLCSIM \citep{pylcsim}. For simplicity we assume a simple powerlaw shaped power spectral density (PSD). Meanwhile, the background count rate (within the source region) is assumed to be intrinsically invariable and set at 0.03 counts/s within 3--79 keV for each NuSTAR module (FPMA and FPMB), which is a typical value of the sources analyzed in \citet{Kang_2022}. The input parameters in this step include count rate ($\rm CR$), fractional root mean square ($\rm frms$, within the exposure time) and index of the power spectra ($\alpha$). Then we add independent Poisson fluctuations to the source and background light curves for FPMA and FPMB respectively. Here we obtain three light curves for each module, including the source light curve $\rm LC_{\rm source}$, the light curve of the background within the source region $\rm LC_{\rm bkg}^{\rm src~reg}$ and the light curve of background measured within the background region $\rm LC_{\rm bkg}^{\rm bkg~reg}$. The net light curves (FPMA + FPMB) are hence calculated as following. 
\begin{equation}\label{eq1}
\rm LC_{\rm net} = \sum_{A,B} LC_{\rm source} +  LC_{\rm bkg}^{\rm src~reg} - LC_{\rm bkg}^{\rm bkg~reg}*f_{scale}
\end{equation}

The simulated light curves are then split horizontally into two flux levels (high versus low) with a single cut in count rate. The critical count rate is chosen to make the high and low states possess similar total net counts.

\par
\begin{figure}
\centering
\subfloat{\includegraphics[width=0.5\textwidth]{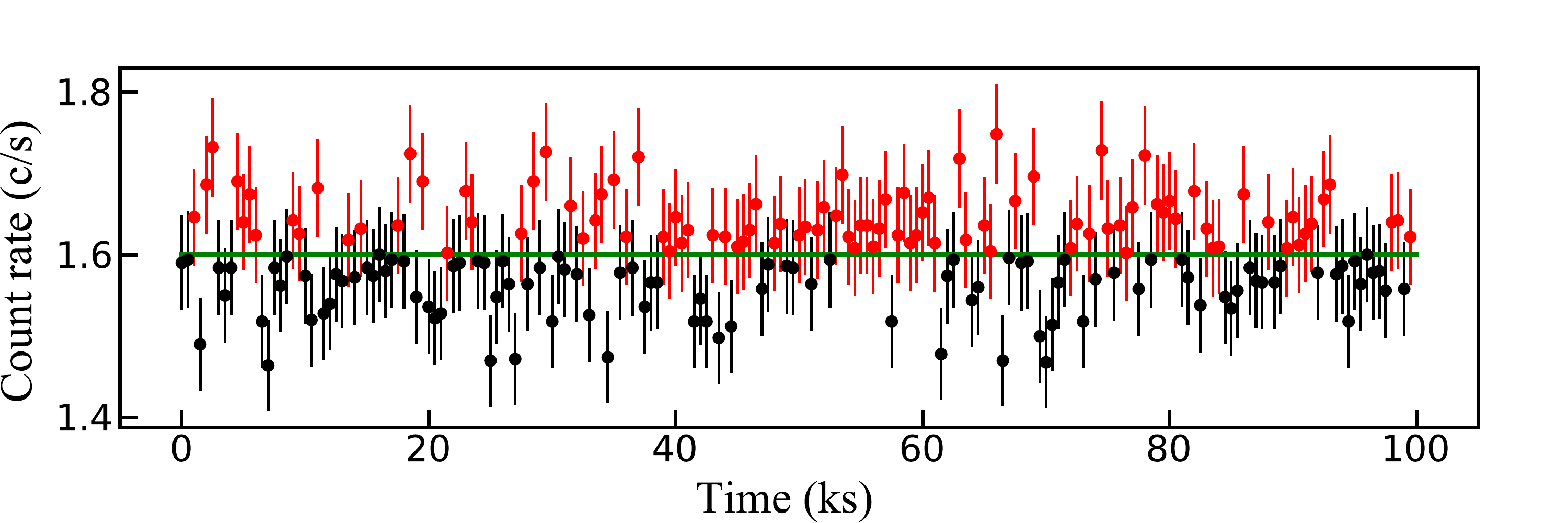}}\\
\vspace{-10mm}
\subfloat{\includegraphics[width=0.5\textwidth]{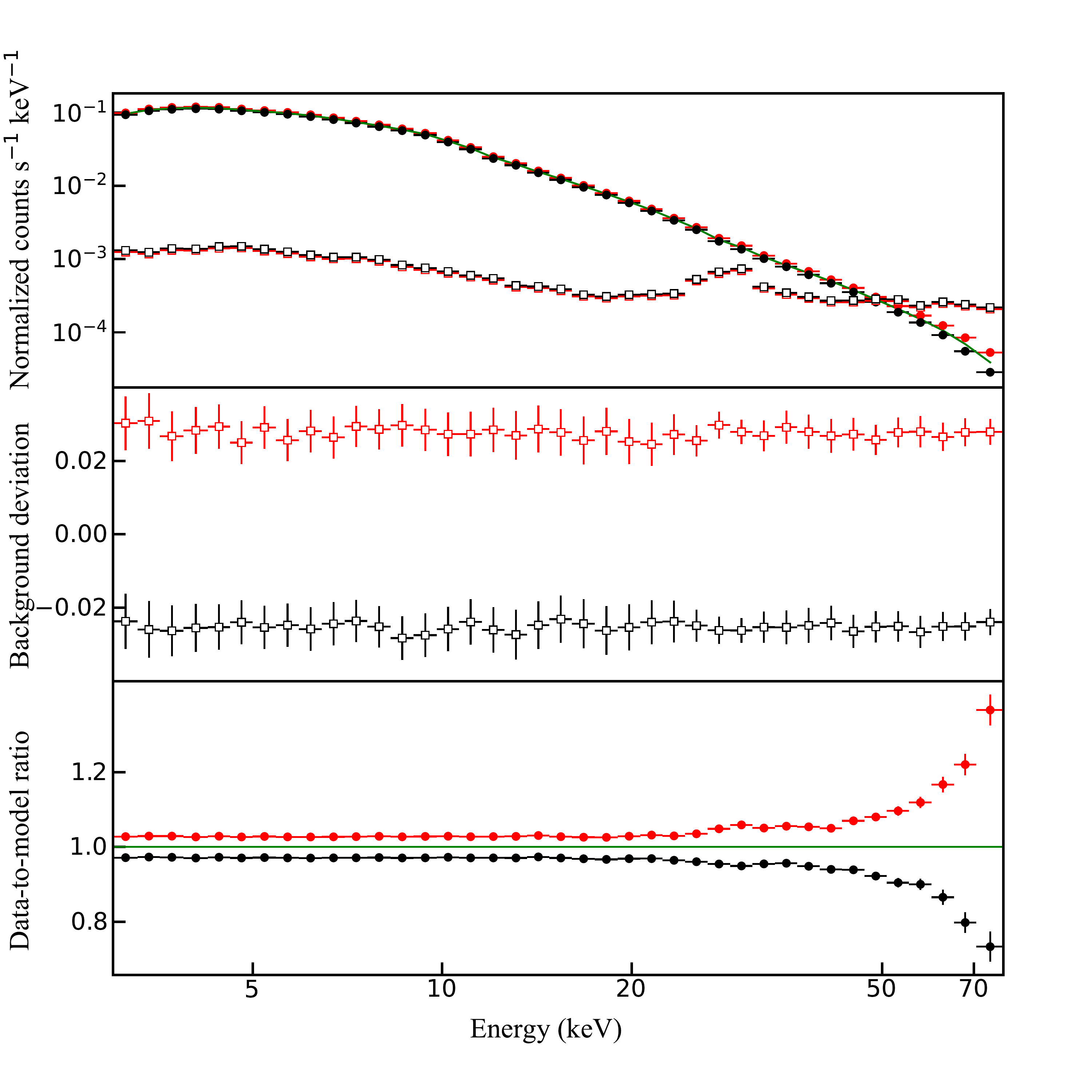}}
\caption{\label{fig:bias} Top panel: a 100 ks section of the total 50 Ms simulated light curve (FPMA+FPMB) of an intrinsically invariable source. The light curve (with a bin size of 500 s) is split into high (red) and low (black) states with a single cut in count rate. The 2nd panel: the input spectral model (green line, a simple powerlaw with $\Gamma$ = 1.85) and the simulated spectrum $\rm S_{net}$ (red and black solid circles) of the high and low states, respectively. The corresponding simulated background spectra $\rm S_{bkg}^{bkg~reg}*f_{scale}$(open squares) are over-plotted for comparison. The 3rd panel: the difference between the simulated $\rm S_{bkg}^{src~reg}$ and $\rm S_{bkg}^{bkg~reg}*f_{scale}$, normalized by the input background spectrum, for an enlarged view of the difference between the high and low state background spectra. The bottom panel: the data-to-model ratio between the mock spectra and the input model.}
\end{figure}

 With \Nu response files, we use XSPEC \citep{Arnaud_1996} to derive the count distribution of a certain spectral model. We adopt the typical background spectrum of  \Nu observations, estimated by NUSKYBGD \citep{Wik_2014} to obtain the spectral distribution of background photons (Figure \ref{fig:bias}). To simulate source spectra and spectral distribution of source photons, we employ the response files (RMF and ARF) of the NuSTAR observation 60301004002 \citep{Barua2021} produced using NuSTARDAS following the standard pipeline. For each light curve, the simulated counts are randomly cast into \Nu channels according to the expected count distribution. The mock spectra include the integrated source and background spectrum $\rm S_{tot}$ (= $\rm S_{src}$ + $\rm S_{bkg}^{\rm src~reg}$) within $\rm Area_{src}$, and the background spectrum $\rm S_{bkg}$ within $\rm Area_{bkg}$, for FPMA/FPMB, and high/low states, respectively. The simulated net spectrum $\rm S_{net}$ is then  $\rm S_{src}$ + $\rm S_{bkg}^{src~reg}$ - $\rm S_{bkg}^{bkg~reg}*f_{scale}$. FPMA and FPMB spectra are then fitted simultaneously, with a cross-normalization difference typically less than 5 percent \citep{Madsen_2015}. 

\section{The hidden biases}

\par To illustrate the concepts of these biases, we first simulate a 50 Ms NuSTAR exposure of a single powerlaw spectrum with photon index $\Gamma=1.85 $, following the procedure in Sec. \ref{sec:method}. The source is assumed to be intrinsically invariable, with a count rate $\rm CR=0.8\,c/s$ (for each \Nu module, hereafter the same) and a light curve bin size of 500 s. The source count rate of 0.8 c/s is also typical among bright AGNs observed with NuSTAR \citep{Kang_2022}. Obviously, splitting a light curve of an intrinsically invariable source is scientifically meaningless. Here it is adopted just to demonstrate the concept of the biases in the most extreme cases. By splitting such a simulated light curve horizontally into two states (high vs low), we clearly see that the source is brighter in the high state than in the low state, solely because of the fluctuations  (Bias\uppercase\expandafter{\romannumeral1 }, see the upper panel in Figure \ref{fig:bias}). 

\begin{figure}
\centering
\subfloat{\includegraphics[width=0.5\textwidth]{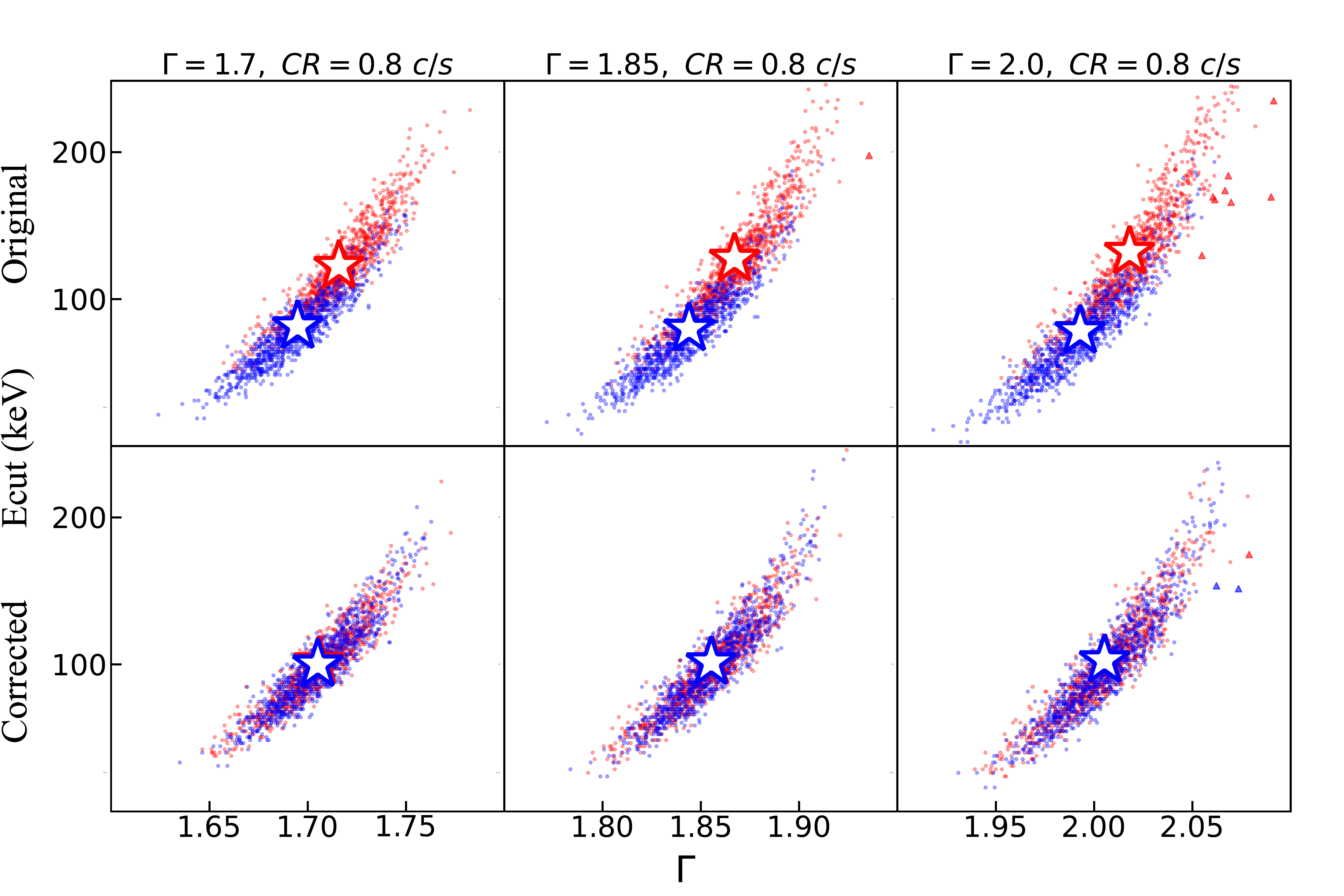}}\\
\subfloat{\includegraphics[width=0.5\textwidth]{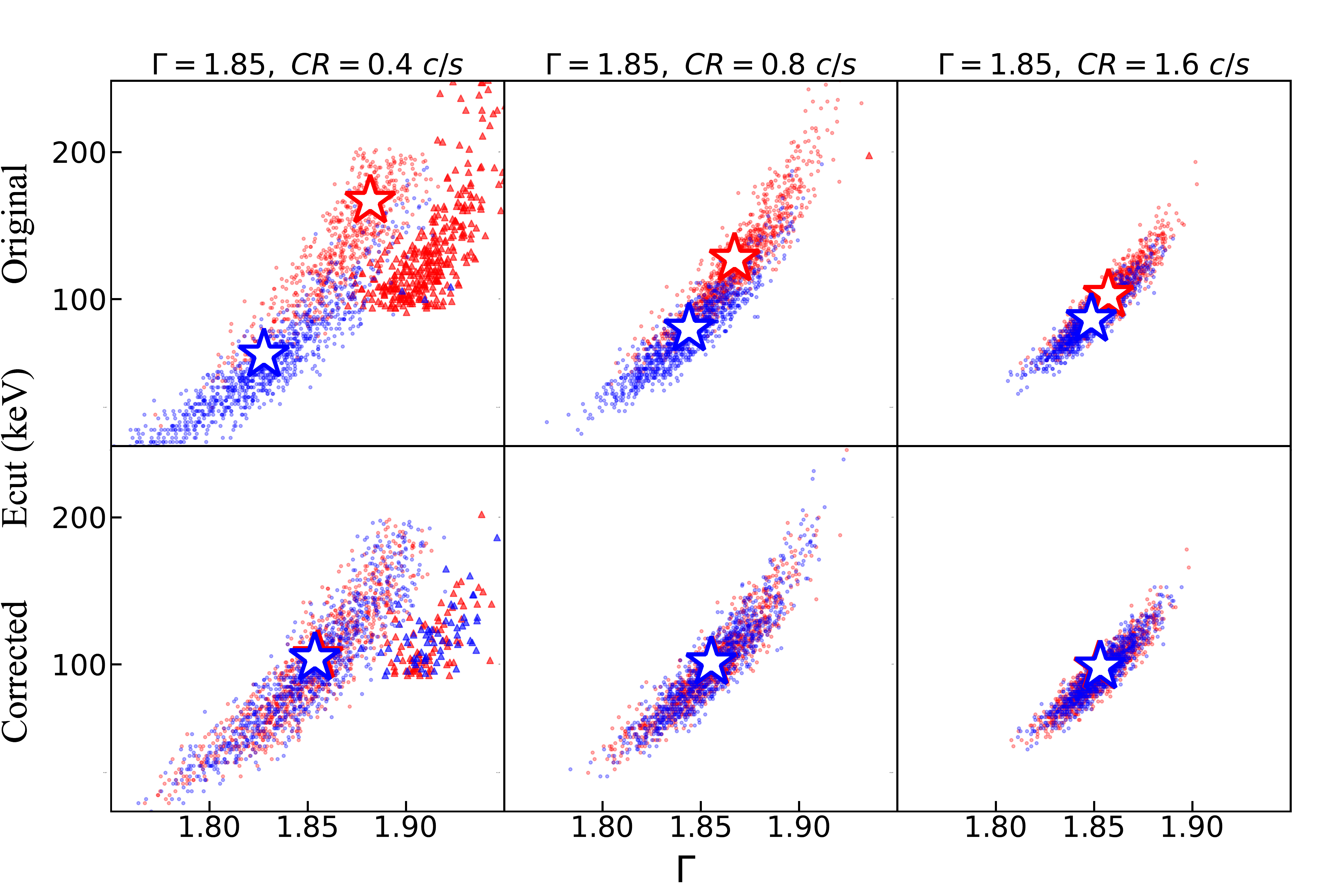}}
\caption{\label{fig:contour_inva}. The output \ec versus $\Gamma$ from the 1000 simulations, of the high and low states (red vs blue), in the case of an intrinsically invariable source with different photon index $\Gamma$ (upper panel, from left to right, $\Gamma$ = 1.7, 1.85, and 2.0) and count rate (lower panel, from left to right, CR = 0.4, 0.8, 1.6 ct/s). The results before and after applying our bias corrections are marked as ``original" and ``corrected", respectively. Triangles indicate lower limits to \ec in some simulated exposures where \ec can not be well constrained. Open stars are the mean values derived with the Kaplan-Meier estimator (to treat the lower limits). Note in some sub-panels, the blue and red stars completely overlap and only the blue star is visible. }
\end{figure}  

\begin{figure*}
\centering
\subfloat{\includegraphics[width=1.0\textwidth]{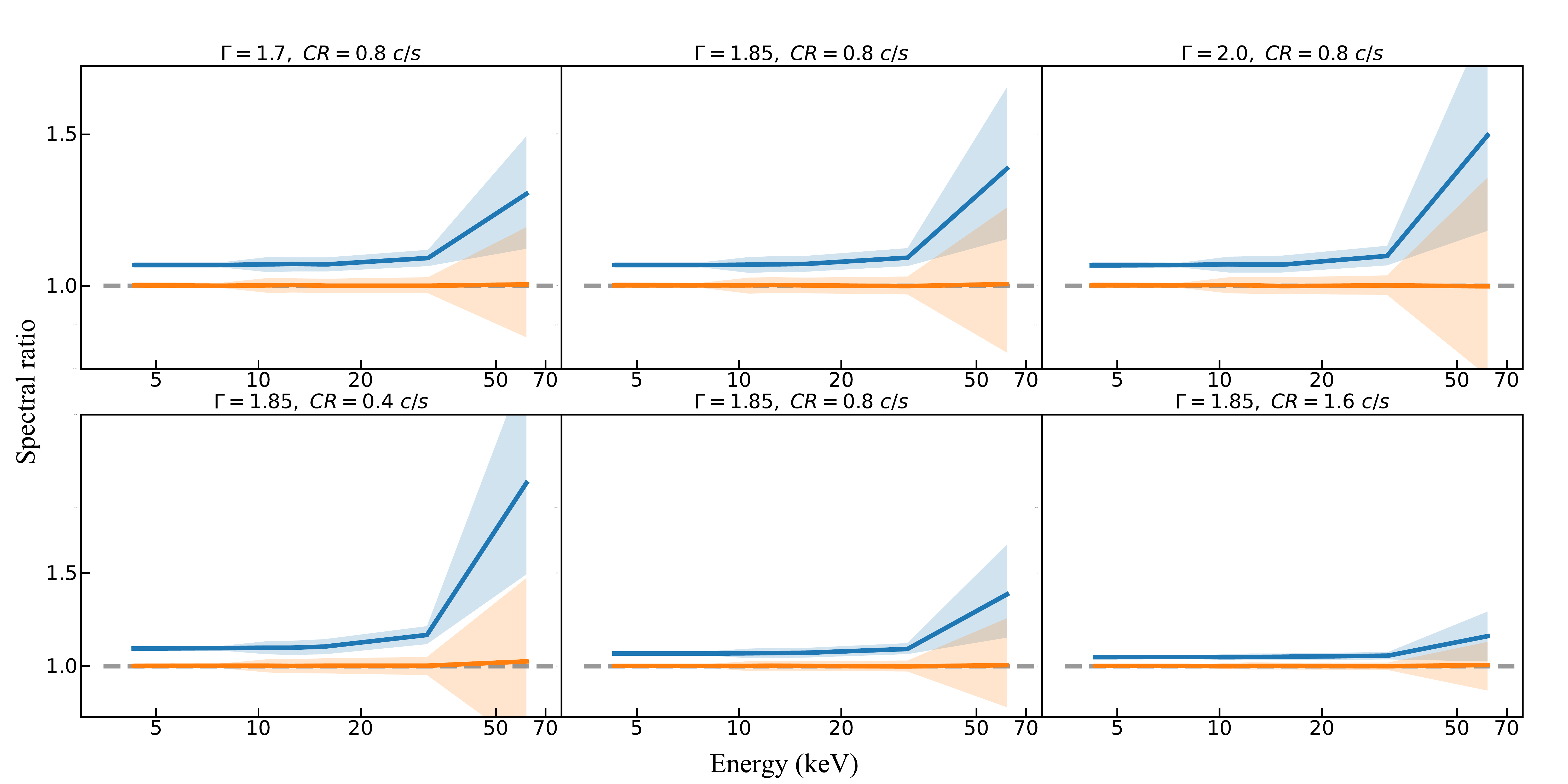}}
\caption{\label{fig:invariable} The spectral ratios (high over low) in the case of an invariable source with different photon index $\Gamma$ (upper panel) and count rate CR (lower panel). The grey dashed line stands for the input spectral model ratio, while the blue/orange solid line for the spectral ratios of the simulated spectra before/after correction respectively. Shadows indicate the 16--84 quantiles of the 1000 simulations. }
\end{figure*}  

\par We present the simulated net spectra (FPMA) and corresponding background spectra in the 2nd panel of Figure \ref{fig:bias}. To better demonstrate the effect of BiasII, in the 3rd panel we plot the difference between the simulated $\rm S_{bkg}^{src~reg}$ and $\rm S_{bkg}^{bkg~reg}*f_{Scale}$, normalized by the input background spectrum, for the high and low states, respectively. It could be seen that, the canonical background estimation ($\rm S_{bkg}^{bkg~reg}*f_{Scale}$) under-estimates the background spectrum for the high state, and vice versa for the low state. In the lower panel of Figure \ref{fig:bias} we plot the ratio of the simulated net spectra to the input model for the high and low states, respectively. At lower energies where background counts are negligible, BiasI is dominant, simply yielding a larger (smaller) flux normalization for the high (low) state spectrum without altering the spectral slope. As the background fraction in the total spectrum generally increases with energy (the background spectrum is harder than the source spectrum), at higher energies the effect of biased background subtraction due to  BiasII emerges. This yields a harder net spectrum in the high state with the data to input model ratio upwarping at high energies. Contrarily, the low state spectrum exhibits downwarping at high energies in the data to input model ratio plot. Apparently, such warping at high energies may bias the measurement of spectral parameters\footnote{Note in an extreme case that the background spectrum has a shape identical to that of the source spectrum, BiasII degenerates into BiasI, i.e., only biases the measurement of flux normalization without altering the spectral shape.}, particularly $E_{\rm cut}$/$T_{\rm c}$ which is rather sensitive to the spectra at highest energies.

\par To further quantify the effect of BiasII on spectral parameter (including $E_{\rm cut}$ and $\Gamma$) measurements, in the following simulations we adopt a cutoff powerlaw (cutoffpl in XSPEC) as the input model (with $E_{\rm cut}$ set at 100 keV). We simulate 100 ks long NuSTAR exposures (typical for NuSTAR observations of AGNs), bin the light curves with 500 s, and split it into high and low states with a single count rate cut. To better demonstrate the effects of the biases, we simulate 1000 groups of 100 ks exposures (to reduce the effect of significant photon noises in a single 100 ks long exposure), perform spectral fitting (3--79 keV) to each individual exposure, and present the mean output parameters of the 1000 exposures. We adopt $\chi^{2}$ statistics and calculate the errors/limits following the $\Delta \chi^{2}= 2.71$ criterion, corresponding to the 90\% confidence level. Individual spectra are rebinned to achieve a minimum of 50 count bin$^{-1}$ using GRPPHA, except that the highest energy bin below 79 keV is manually set to include all further channels up to 79 keV. That is to say, if the total counts of the last few to-be-rebinned channels below 79 keV are less than 50, we manually re-bin them into the existing highest energy bin $<$ 79 keV, to ensure all the simulated spectra are fitted in the same energy range of 3 -- 79 keV. \ec and $\Gamma$ are measured for each individual exposure. Since in some individual exposures \ec could be statistically undetectable and only lower limits could be obtained, we obtain the mean of \ec from 1000 groups of simulations using Kaplan-Meier method within the survival analysis package ASURV \citep{Feigelson_1985}. Note during simulations we assure the spectra of high and low states having similar total net counts and thus similar signal-to-noise ratios. In this way, the mean \ec of the two states could be directly compared without being biased by the different spectral S/N \citep{Kang_2022}. Besides presenting the mean output spectral parameters, we also obtain the ratio of high to low state spectra (the spectral ratio technique, \citealt{Zhangjx2018, Kang_2021}). The spectra ratio plot is obtained for each individual exposure, and we plot the median spectra ratio from 1000 simulated exposures to directly illustrate the effects of the biases. 

\par In the first place we again assume an intrinsically invariable source. We plot in Figure \ref{fig:contour_inva} the output \ec versus $\Gamma$ of the 1000 simulations for the high and low state, along with the mean \ec and $\Gamma$, for various input spectral slope $\Gamma$ and source net count rates. As expected, the output \ec of the high state are considerably and systematically larger than those of the low state, producing a pseudo \ec variation. Meanwhile, the output $\Gamma$ of the high state are slightly larger than those of the low, likely due to the degeneracy between \ec and $\Gamma$ \citep{Kang_2021}. Therefore, the biases may produce an artificial ``hotter/softer-when-brighter'' variation trend for an invariable source.

\par Meanwhile, the median spectral ratio plots (high state over low state) are shown in Figure \ref{fig:invariable}. The spectra ratio plots clearly illustrate the effect of BiasI and BiasII, that the high state spectra are biased to have artificially larger flux normalization, and clear upwarping at higher energies, similar to the trend shown in the bottom panel of Figure \ref{fig:bias}.

\par We further explore how the effects of such biases change with spectral parameters. The top panels of Figure \ref{fig:contour_inva} and Figure \ref{fig:invariable} show the cases of different $\Gamma$. As the $\Gamma$ increases, the background fraction at the high energy end increases (for given 3--79 keV net count rate), boosting the effect of BiasII, while the effect of \first (source fluctuations) is barely influenced. The bottom panels of Figure \ref{fig:contour_inva} and Figure \ref{fig:invariable} show the cases of different input source net count rate CR. As expected, the effects of both biases are weaker when the source is brighter. 

\par
\begin{figure}
\centering
\subfloat{\includegraphics[width=0.5\textwidth]{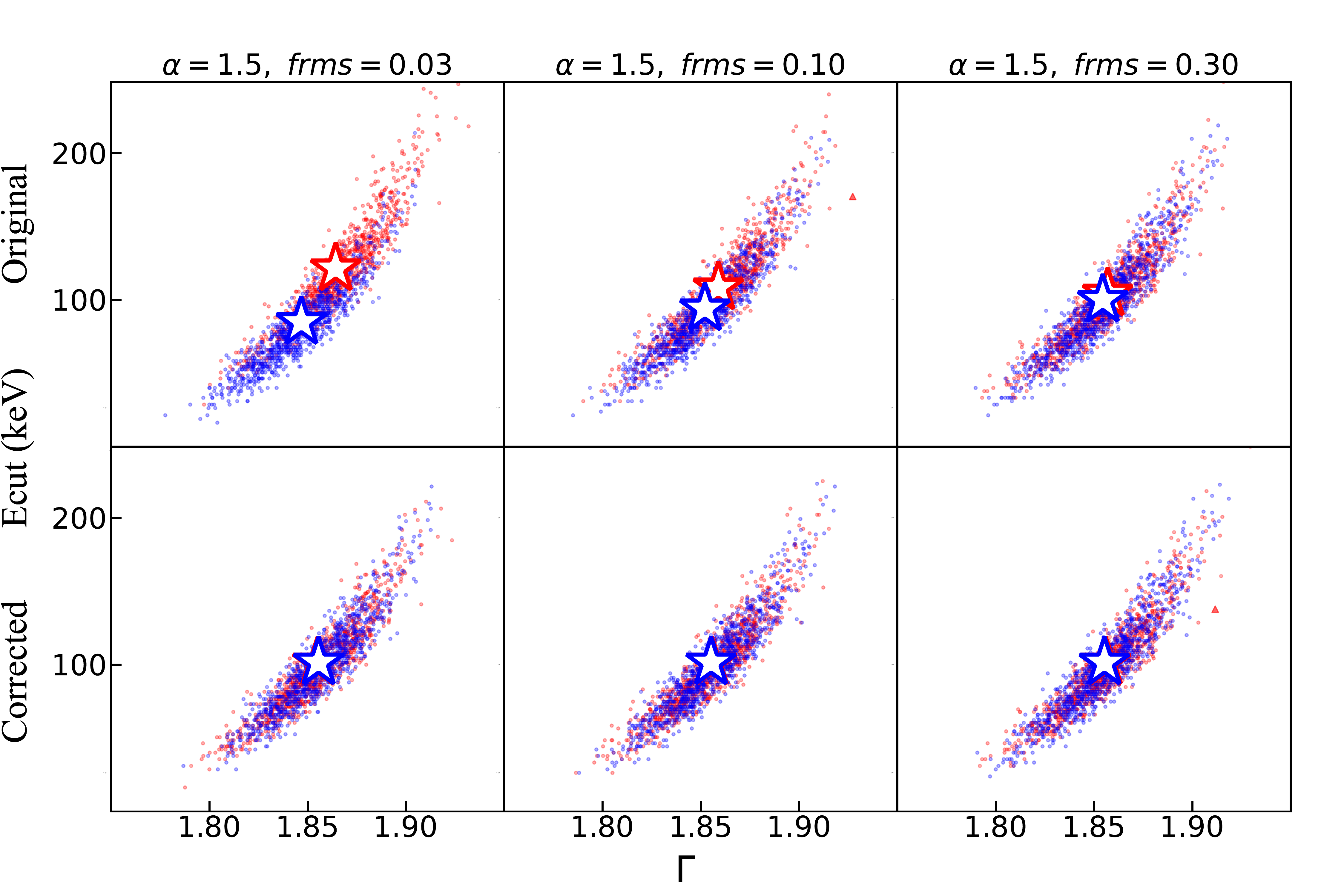}}\\
\subfloat{\includegraphics[width=0.5\textwidth]{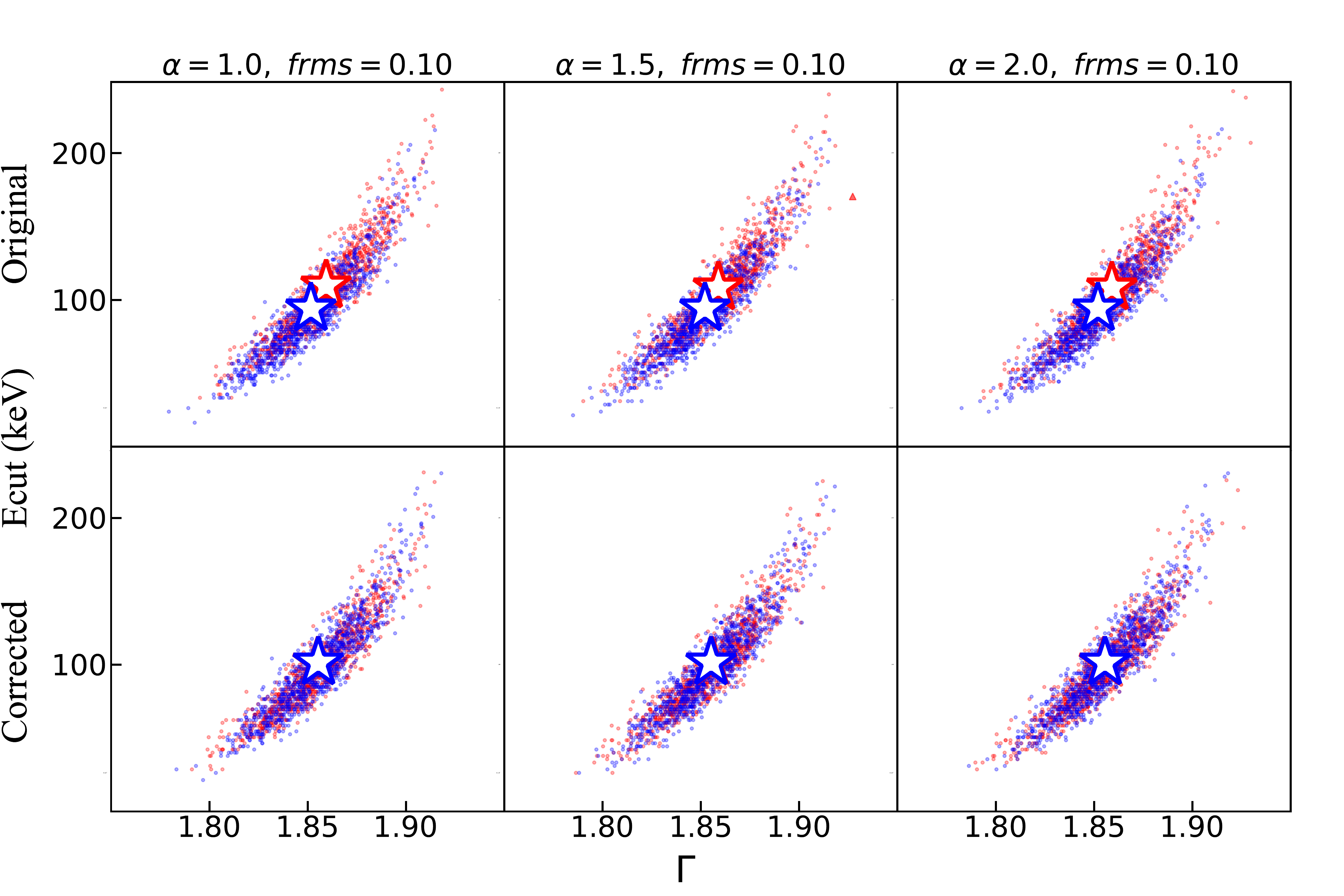}}
\caption{\label{fig:contour_va}. Similar to Figure \ref{fig:contour_inva} but for an intrinsically variable source ($\rm CR=0.8\,c/s$ and $\Gamma = 1.85$), with different fractional root-mean-squares (frms, upper panel) and indices of the power spectral density ($\alpha$, lower panel). 
}
\end{figure}

\begin{figure*}
\centering
\subfloat{\includegraphics[width=1.0\textwidth]{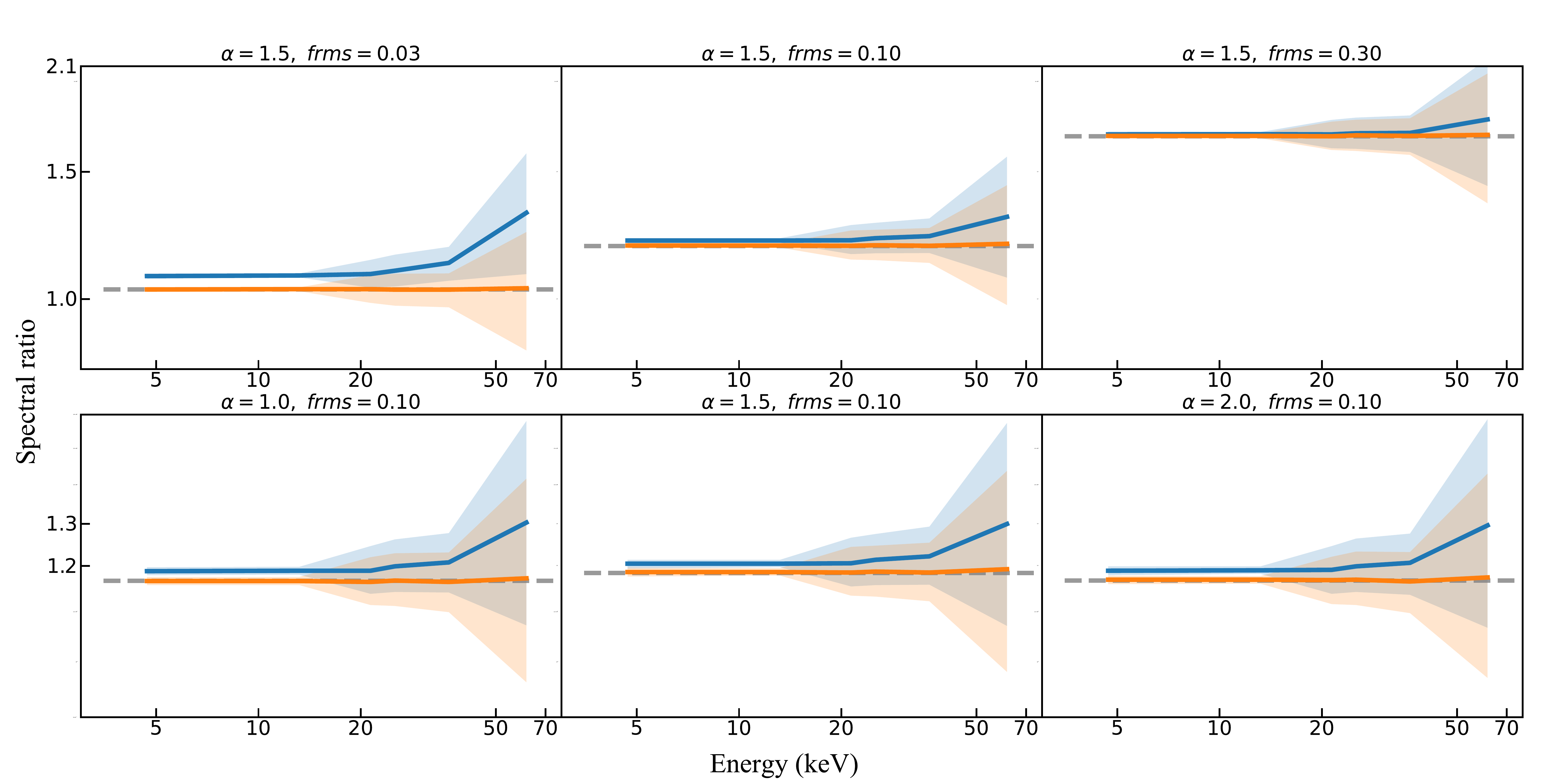}}
\caption{\label{fig:variable}
Similar to Figure \ref{fig:invariable} but for an intrinsically variable source ($\rm CR=0.8\,c/s$ and $\Gamma = 1.85$),
with different fractional root-mean-squares (frms, upper panel) and indices of the power spectral density ($\alpha$, lower panel). 
}
\end{figure*}

 Next we inspect a more practical scenario where the source is intrinsically variable. For simplicity we assume the variable source has a fixed spectral shape. 
The source is assumed to have a $\Gamma=1.85$, an \ec $=100$ keV, and a mean $\rm CR=0.8\,c/s$. Using pyLCSIM, we investigate cases of different fractional root-mean-squares (frms = 0.03, 0.1 and 0.3, respectively) and indices of the power spectral density ($\alpha$ = 1.0, 1.5 and 2.0). Note the equivalent frms induced by the source (background) photon Poisson fluctuations are $\sim $ 120, 35 and 10 percent ($\sim $ 24, 7 and 2 percent) of the intrinsic frms when the intrinsic frms are 0.03, 0.1 and 0.3 respectively. As shown in Figure \ref{fig:contour_va} and Figure \ref{fig:variable}, the effects of both biases become weaker when the source is intrinsically more variable, while the slope of the PSD (the time scale of the variation) barely influences the effects of the biases. Note, while the effects of BiasI and BiasII appears rather weak when the intrinsic variation of the source is strong (see the upper right panel of Figure \ref{fig:variable}), they are still appreciable and non-negligible compared with statistical noise, in scenario where the intrinsic variation amplitude is weak (see other panels in Figure \ref{fig:variable}). 

\section{Assessments and corrections for the biases}      
\par Clearly, the strength of the hidden biases we observe depends on various factors. For BiasI, the factors include the brightness of the source, the amplitude of the intrinsic variation, and the bin size of light curves used to split it into hard/low states. For BiasII, additional factors includes the background count rate, the area of the background region, the spectral shape of the background spectrum and of the source spectrum. The effect of light curve bin size, which we have not discussed previously, is straightforward that larger bin size could reduce the Poisson fluctuations in each bin thus reduce the biases aforementioned, but at the expense of smearing out rapid variations.

\par The effect of the biases also depends on the exact variation pattern of the light curve. For instance, for a strictly monotonically increasing/decreasing light curve, splitting it horizontally into high/low states with a count rate cut is effectively equivalent to splitting it vertically, which means there exist no biases in the approach.  

\par
\begin{figure}
\centering
\subfloat{\includegraphics[width=0.5\textwidth]{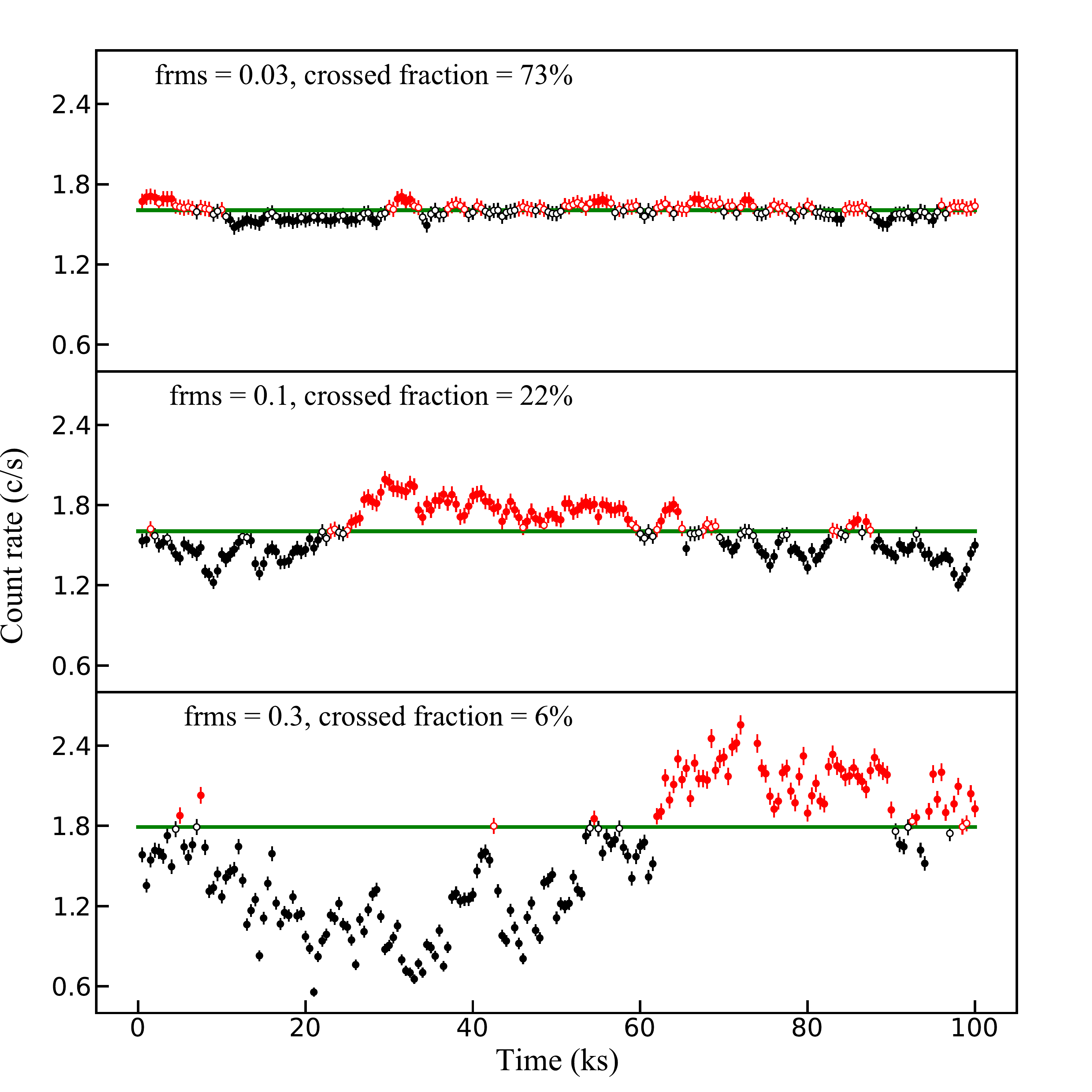}}
\caption{\label{fig:lcs} The example light curves used as input for Figure \ref{fig:variable}. The open points represent the bins crossed by the count rate cut within the 1$\sigma$ error range. The biases are expected to be weaker in case of smaller ``crossed fraction".}
\end{figure}

 It is often hard to evaluate the effects of the biases in various exposures with distinct light curves. Here we first provide a simple principle to roughly assess the effect of the hidden biases for a single light curve, by which we might tell at a glance if the biases are strong in the interested case and elaborating corrections are indispensable. In Figure \ref{fig:lcs} we plot the simulated light curves with different variation amplitudes, which are used as inputs in Figure \ref{fig:variable}. The open points represent the bins crossed by the count rate cut within the 1$\sigma$ error range\footnote{Here, using different confidence ranges such as 2$\sigma$ or 3$\sigma$ also works.}. Note the origin of the biases is that, after fluctuations, some light curve bins may have been ``mistakenly" classified into high or low states due to fluctuations. Therefore, this ``crossed fraction'' can somehow manifest the strength of the biases, the smaller the ``crossed fraction'', the weaker the biases. 

\par Can one simply drop the bins crossed by the count rate cut (within a certain error range) to avoid the biases? The answer is no. As long as the count rate criteria is set, the two biases emerge. Dropping the crossed bins effectively slightly increases the count rate cut for high state, and decreases it for the low state, so the cuts and biases still exist. Similarly, in the scenario where the light curve is split horizontally into more than two flux states, the two biases still exist between various states. 

\par
\begin{figure}
\centering
\subfloat{\includegraphics[width=0.5\textwidth]{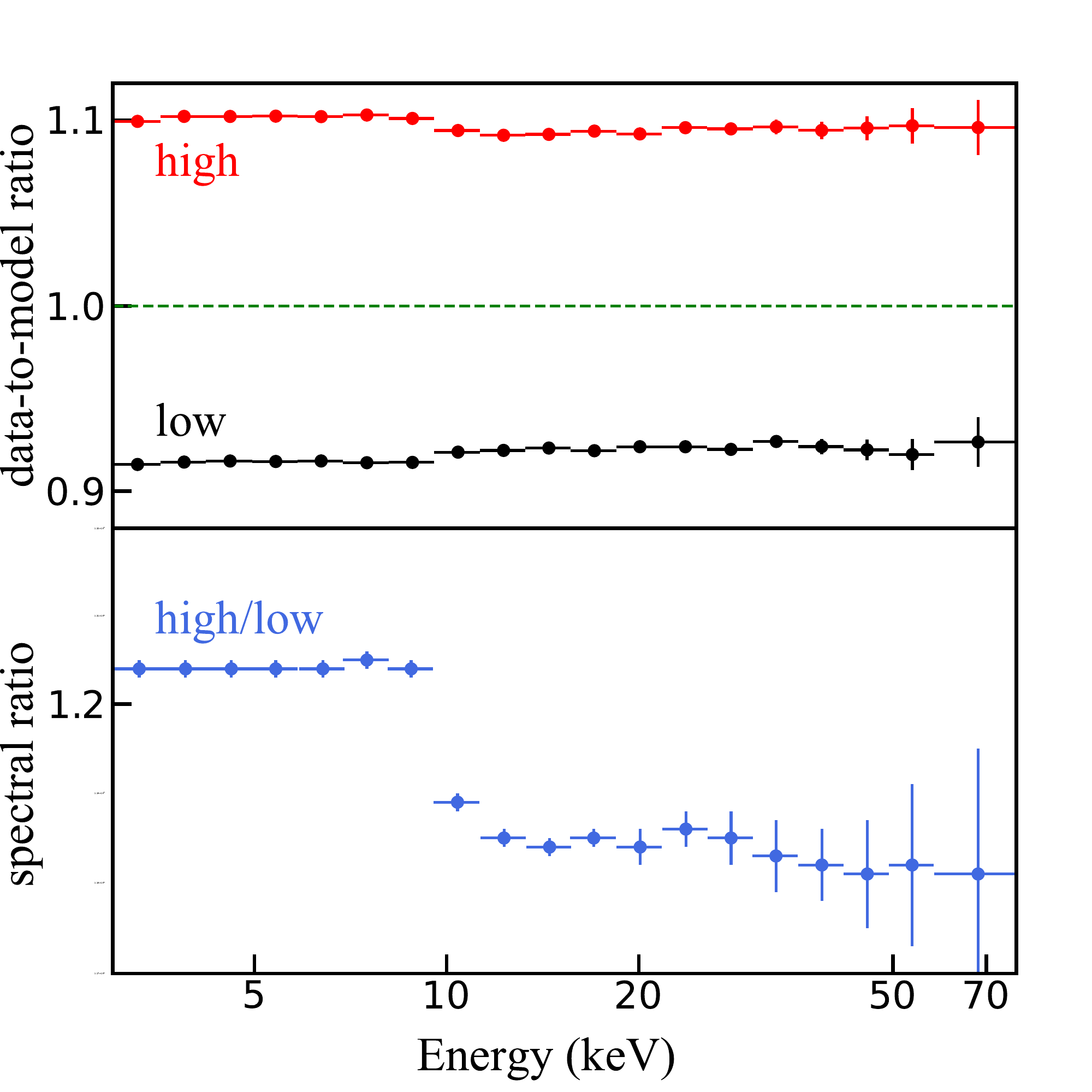}}
\caption{\label{fig:fracture} The ratios of the output spectra (high and low states) to the input average model (upper panel) and the spectral ratio between the high and low states (low panel), when using low energy (3--10 keV) NuSTAR light curves for flux division.
}
\end{figure}

 Can one use the low energy (e.g., 3--10 keV) light curve, where the background counts are negligible, to split the exposure horizontally, but analyze the whole band (3--79 keV for NuSTAR) spectra, to avoid BiasII? In this scenario, the independent photon fluctuations in different energy bands will introduce extra bias. Figure \ref{fig:fracture} shows the case where a 50 Ms exposure of a source with frms = 0.1 (a single powerlaw with $\Gamma = 1.85$) is split into high and low states based on 3--10 keV light curves. While \second could indeed be disposed, the whole band spectra are broken and twisted in this case (because only the 3--10 keV spectra in low/high states are biased). Thus, in principle, one should use the same band to split the light curve and analyze the spectra\footnote{Alternatively, one may split the light curve in one band, but analyze the spectra in a totally independent band. While this option is free from the biases we introduced assuming the photon fluctuations at various energies are completely independent, the loss of bandwidth in spectral fitting is unacceptable in most studies.}.

\par In \S\ref{sec:method} we simulate photon fluctuations in the net light curve following Equation \ref{eq1}, which is the root cause of the biases discussed in this work.
The same equation also enlightens the way to correct the biases. If we know the intrinsic light curve, we could repeat the first part of simulation processes in \S\ref{sec:method}, i.e., simulating mock light curves by adding photon fluctuations to the input one, split the mock light curves into low/high states, and compare the output low/high states with the input ones to derive the correction factors for biases. Nevertheless, practically the only available light curves are the observed ones, already suffered source and background photon fluctuations. Taking the observed light curves as the intrinsic ones and calculating the correction factors will lead to insufficient correction, because the observed light curves have larger frms than the intrinsic ones, and larger frms means weaker biases (see Figure \ref{fig:variable}). Therefore, the correction factors should be derived with two steps: estimating the intrinsic light curve, and performing simulations to obtain the correction factors. 

\par Similar to Equation \ref{eq1}, the observed $\sigma^2_{\rm rms}$ of a light curve could be described as:
\begin{footnotesize}
\begin{equation} \label{eqsigma}
\rm \sigma_{rms~obs}^2 =  \sigma_{rms~int}^2 + \sigma^2_{src} + f_{scale} *(1 + f_{scale}) * \sigma_{bkg}^2   
\end{equation}
\end{footnotesize}
where $\rm \sigma^2_{rms~int}$ describes the intrinsic variability of the source count rate; $\rm \sigma^2_{src}$ the mean variance of source count rate in each light curve bin due to Poisson fluctuation, $\rm \sigma^2_{bkg}$ the mean variance of  background count rate (due to Poisson fluctuation) measured in background region, and 
f$_{\rm scale}$ the ratio of the areas of $\rm Area_{src}/Area_{bkg}$. This equation takes into account the effect of photon fluctuations of both source and background counts (akin to convolving of the intrinsic count rate distribution with the probability distribution due to Poisson fluctuations).\footnote{Note here for simplicity we assume the count rate distribution of the light curves is Gaussian, and the number of counts within each light curve bin is large enough to be treated as Gaussian variables.} 

\par Thus for each observed light curve, we could easily obtain its $\rm \sigma^2_{rms~int}$ with Equation \ref{eqsigma}.
We could then regulate the observed net light curve following
\begin{footnotesize}
\begin{equation} \label{regulate}
\rm LC_{obs}^{'} = [LC_{obs} - mean (LC_{obs})]*\sigma_{rms~int}/\sigma_{rms~obs} + mean (LC_{obs})
\end{equation}
\end{footnotesize}
The regulated light curve $\rm LC_{obs}^{'}$ will have the same mean count rate as the observed one, the same variation pattern, but the intrinsic variance $\rm \sigma^2_{rms~int}.$\footnote{
If the derived $\rm \sigma_{rms~int}^2 \leq 0$, an even light curve is adopted as the intrinsic one by setting $\rm \sigma_{rms~int}^2$ to 0. }
Using the regulated light curve as an input intrinsic source light curve and adding photon fluctuations (of both source and background photons) to it, we repeat the simulation processes in Sec. \ref{sec:method} to mimic the effects of photon fluctuations that already exist in the observed light curve. 

\par Starting from $\rm LC_{obs}^{'}$ and the observed $\rm LC_{bkg}$, we generate 1000 sets of simulated light curves (including $\rm LC_{\rm src}$, $\rm LC_{\rm bkg}^{\rm src~reg}$, $\rm LC_{\rm bkg}^{\rm bkg~reg}$, and $\rm LC_{\rm net}$, see Equation \ref{eq1}) by adding Poisson fluctuations to input source and background light curves. We then split each simulated $\rm LC_{\rm net}$ into high/low states accordingly with given horizontal count rate cuts, and measure two mean correction factors (averaging over 1000 simulations) for each state:
\begin{footnotesize}
\begin{equation}\label{eq2}
\begin{aligned}
\rm f_1 = \frac{mean\,CR^{output}_{src}}{mean\,CR^{input}_{src}}, 
\rm f_2 = \frac{mean\,CR^{bkg~reg}_{bkg} \cdot f_{scale}}{mean\,CR^{src~reg}_{bkg}} 
\end{aligned}
\end{equation}
\end{footnotesize}
where, for a given state, mean $\rm CR^{src~reg}_{bkg}$ is the average of the simulated background count rate in the source region,
mean $\rm CR^{bkg~reg}_{bkg}$ the average of the simulated background count rate in the background region,
mean CR${\rm ^{input}_{src}}$ the average count rate measured from $\rm LC_{obs}^{'}$,  and CR${\rm ^{output}_{src}}$ the average count rate from the simulated $\rm LC_{\rm src}$, respectively. Then
$\rm f_1$ and $\rm f_2$ could be used to correct the effects of BiasI and BiasII, respectively.
To perform the corrections, we modify the header ``AREASCAL" (the scale of the area) of the source and background spectra files, multiplying that of the source by $\rm f_1$ and that of the background by $\rm f_1 * f_2$. 

\par Applying these corrections to the simulated spectra as described in \S\ref{sec:method}, we find the corrections are highly effective and efficient. As shown in Figure \ref{fig:invariable} \& \ref{fig:variable},
the deviations from the input spectra ratio (blue lines versus grey lines) caused by BiasI and BiasII now almost disappear (see orange lines).

\par Note in Figure \ref{fig:invariable} (for an intrinsically invariable source), the correction for the net flux level is slightly insufficient (with the orange line lying slightly above the grey line). This is as expected since for simulated net light curves of an intrinsically invariable source, the $\rm \sigma_{rms~int}^2$ derived following Equation \ref{eqsigma} may still be positive thus yielding artificial flux variation.

\par Using this correction method, we first revisit the two sources analyzed in \citet{Barua2020, Barua2021}. We find [$f_{1}^{\rm high}, f_{2}^{\rm high}, f_{1}^{\rm low}, f_{2}^{\rm low}$] = [1.006, 0.950, 0.995, 1.029] for Ark 564, and [1.008, 0.981, 0.990, 1.012] for ESO 103-035, respectively. The employed models for the two sources are xillvercp and relxillcp respectively \citep{Garcia_2013, Garc_2014}, with errors corresponding to 1$\sigma$ level. Through fitting the spectra, for Ark 564, we find $kT_{\rm e}^{\rm high} = 20.0^{+4.7}_{-1.8}$ keV and $kT_{\rm e}^{\rm low}>34.5 $ keV before the correction, while $19.2^{+3.8}_{-1.9}$ keV and $>37.0$ keV after the correction. For ESO 103-035, we find $kT_{\rm e}^{\rm high} = 36.0^{+10.1}_{-7.8}$ keV and $kT_{\rm e}^{\rm low} = 20.2^{+4.5}_{-1.1}$ keV before the correction, while $34.7^{+10.0}_{-7.7}$ keV and $21.1^{+4.6}_{-1.2}$ keV after the correction. Therefore, the corrections barely change the best-fit spectral parameters, thus would not alter the scientific results of \citet{Barua2020, Barua2021}. For Ark 564, this is because it is highly variable (with frms = 0.34) thus the effects of the biases are expected to be weak. Meanwhile, the 42 ks NuSTAR light curve of ESO 103-035 (though with lower frms = 0.13) exhibits a (nearly) monotonic decreasing trend (see Figure 1 of \citealt{Barua2021}) for which the effects of the biases are also weak.  

\par  Moreover, we also apply the correction to the study of \citet{Parker2014}, which used flux-resolved spectroscopy to investigate the variation of the reflection fraction R in Mrk 335. The effects of the biases are found negligible in this source. For instance, we compare the reflection fraction R of the model relxill between the very low and very high states of Mrk 335 as defined in \citet{Parker2014}. We find R$_{\rm very~low}$ $>$ 6.83 and R$_{\rm very~high}$ = 1.72$_{-0.64}^{+0.81}$ before applying the correction, while R$_{\rm very~low}$ $>$ 6.91 and R$_{\rm very~high}$ = 1.70$_{-0.63}^{+0.81}$ after the correction.  Thus applying the correction even tinily strengthens the conclusion of \citet{Parker2014}, i.e., larger R in the very low state, while smaller R in the very high state.

\begin{figure}
\centering
\subfloat{\includegraphics[width=0.5\textwidth]{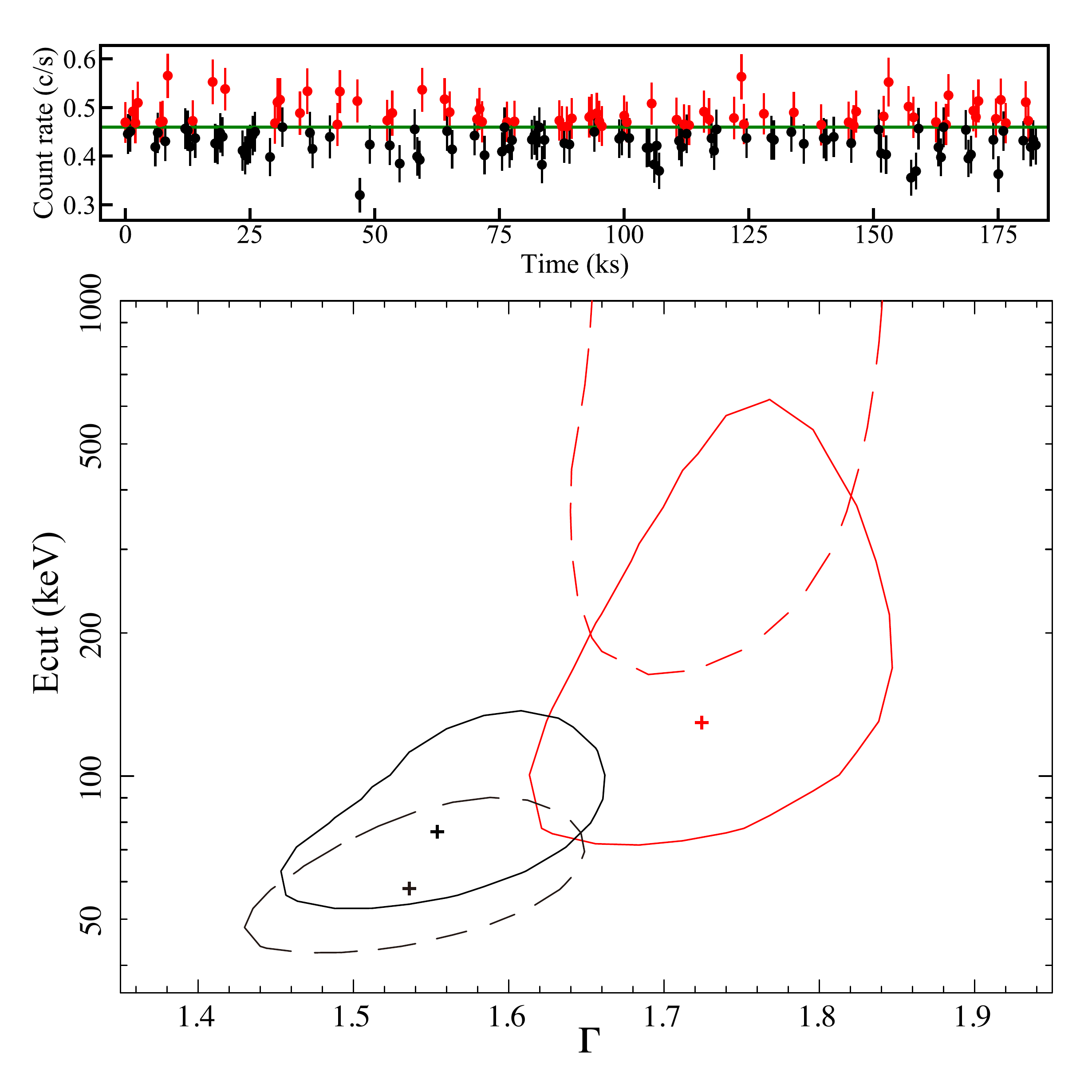}}
\caption{\label{fig:3C109} An example of NuSTAR observation 60301011004 of 3C 109. Upper panel: The net light curves (FPMA+FPMB) with a bin size of 500 s, which is split into high(red) and low (black) states with a single cut (green line) in count rate. Lower panel: The 1$\sigma$ ($\Delta \chi^2=2.3$) $E_{\rm cut}-\Gamma$ contours of the high(red) and low(black) states, before (dashed line) and after (solid line) the corrections. 
}
\end{figure}

\par Nevertheless, for most \Nu observations in \citet{Kang_2022}, the typical frms is $\sim$ 0.1, thus the effects of the biases could be non-negligible. In Figure \ref{fig:3C109} we show the NuSTAR observation 60301011004 of 3C 109 as a real example. In this observation, the source and background count rates of FPMA are 0.23 and 0.02 c/s respectively, and the frms is 0.09.
The crossed fraction (see Figure \ref{fig:lcs} for definition) is 69\%, thus the effects of the biases are expected to be strong in this observation. The correction factors we derived are [$f_{1}^{\rm high}, f_{2}^{\rm high}, f_{1}^{\rm low}, f_{2}^{\rm low}$] = [1.056, 0.868, 0.953, 1.092]. 
The 1$\sigma$ $E_{\rm cut}-\Gamma$ contours \citep[through fitting the spectra with model pexrav, ][]{Magdziarz_1995} of the high and low states, before (dashed line) and after (solid line) the correction are plotted in Figure \ref{fig:3C109}. Before applying our corrections, the best-fit \ec increases from $58^{+18}_{-11}$ keV for the low state to $>$ 240 keV for the high state (the statistical errors and lower limit are at 1$\sigma$ confidence level for one interesting parameter), yielding a ``hotter/softer-when-brighter" variation trend (larger \ec in the brighter state). However, such a pseudo ``hotter/softer-when-brighter" trend vanishes after correcting the biases (\ec = $77^{+32}_{-18}$/$130^{+147}_{-47}$ keV for the low/high states, thus no clear evidence of \ec variation is detected).

\par Also taking this NuSTAR exposure of 3C 109 as an example, we demonstrate how a smaller f$_{\rm scale}$ could help to reduce the effect of BiasII. Adopting a larger background region 8 times of the source region (f$_{\rm scale}$ = 0.125 intead of 1.0), the derived correction factors to BiasII ([$f_{2}^{\rm  high}, f_{2}^{\rm low}$]) change from [0.868, 1.092] to [0.921, 1.046] (the closer to 1.0 of the correction factor, the weaker the BiasII). As expected, using a much larger background region does moderately reduce the effect of BiasII. But note that such a larger background region is often infeasible for NuSTAR exposures considering the limited available source free region on the detectors and the spatially variable background \citep{Wik_2014}. We note again that even with an infinitely large background region to measure the intrinsic background level, BiasII can not be erased due to the Poisson fluctuations of background photons in the source region.

\par Finally, we note the biases in flux-resolved X-ray spectroscopy we introduced in this work, and the corrections we developed, are not confined to analyses of NuSTAR observations or to AGN studies. They could be applicable to other instruments and various types of X-ray variable sources (X-ray binaries, blazars, etc). When the Poisson fluctuations of photon counts are non-negligible compared with the intrinsic variation, it is always advisable to take into account the two biases, which can be easily corrected with the approach presented in this work (the code is available at \url{https://github.com/USTCKang/corrfrxs}).

\section*{Acknowledgements}
We thank the anonymous referee for his/her highly constructive comments that significantly improved the paper. This research has made use of the NuSTAR Data Analysis Software (NuSTARDAS) jointly developed by the ASI Science Data Center (ASDC, Italy) and the California Institute of Technology (USA). The work is supported by National Natural Science Foundation of China (grants No. 11890693, 12033006 $\&$ 12192221). The authors gratefully acknowledge the support of Cyrus Chun Ying Tang Foundations.

\section*{Data Availability}
The data underlying this article will be shared on reasonable request to the corresponding author.

\bibliographystyle{mnras}
\bibliography{Two_biases}{}

\bsp	
\label{lastpage}
\end{document}